\documentclass[twoside,slac_one]{revtex4}
\usepackage{graphicx}
\usepackage{fancyhdr}
\usepackage{amsmath} % American Mathematics Society standards
\usepackage{bm}% bold math
\usepackage{amsxtra}
\usepackage{amssymb}
\usepackage{amsthm}
\usepackage{latexsym}
\usepackage{lscape}

\def\met{\slash{\!\!\!{\rm E}}_T}
\begin{document}

\title{Tevatron Top Quark Forward-Backward Asymmetry -- \\
Implications for Same-sign Top Quark Pair Production}


\author{Edmond L. Berger}
\affiliation{High Energy Physics Division, Argonne National Laboratory, Argonne, IL, USA}

\begin{abstract}
The forward-backward asymmetry for top quarks measured in proton-antiproton collisions at the Tevatron shows an interesting deviation from standard model expectations.  Among possible interpretations, the exchange of a non-universal, flavor-changing $Z^\prime$ is of some interest as it naturally predicts a top quark in the forward region of rapidity.   To reproduce the size of the Tevatron asymmetry, the couplings of the $Z^\prime$ to standard model quarks must be large,
inevitably leading to production of same-sign top quark pairs at the Tevatron and at the Large Hadron Collider (LHC).  We discuss the constraints on this model from (a) the Tevatron $t \bar{t}$ 
cross section, (b) the Tevatron $t \bar{t}$ invariant mass distribution, and the limits at the Tevatron on 
the same sign top quark pair cross section.  We explore the discovery potential for $tt$ and $ttj$ production in early LHC experiments at 7 TeV and conclude that if a $tt$ signal is not observed with 1~fb$^{-1}$ of integrated luminosity, then a non-universal $Z^\prime$ alone cannot explain the Tevatron forward-backward asymmetry.  Limits on the same sign cross section at the LHC from the CMS collaboration already disfavor this interpretation of the reported asymmetry.  

\end{abstract}

\maketitle

\thispagestyle{fancy}


%%%%%%%%%%%%%%%%%%%%%%%%%%%%%%%%%%
\section{Introduction}
Top quarks are observed to be produced preferentially in the forward hemisphere at the Fermilab Tevatron proton-antiproton collider,  where forward is defined by the direction of the incident proton beam~\cite{Aaltonen:2011kc,Abazov:2011rq}.   A  forward-backward asymmetry in rapidity $A_{\rm FB}$ is predicted in the standard model from higher order QCD contributions~\cite{Kuhn:1998jr}, but the data exceed these expectations by a few standard deviations.  The methods of analysis are 
somewhat different in the two collaborations, and they report different dependence of  $A_{\rm FB}$ on the invariant mass of the $t \bar{t}$ system, and on the rapidity of the top quark, but they agree generally that the asymmetry is larger than SM expectations.  The transverse momentum spectrum of the $t \bar{t}$ system is reported to be softer in the data than in the MC@NLO event generator~\cite{Frixione:2003ei}  used to model the event distributions in Ref.~\cite{Abazov:2011rq}.  The MC@NLO code includes the predicted QCD asymmetry so this discrepancy is worth understanding.    

\begin{figure}
\centering
\includegraphics[clip,scale=0.5]{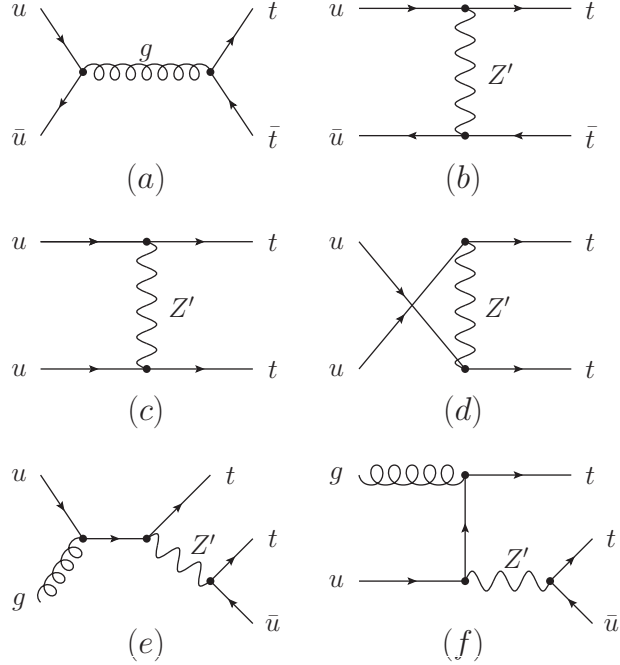}
\caption{Diagrams for (a) $t\bar{t}$ production in the SM, (b) $t\bar{t}$ production
induced by $Z^\prime$ exchange, (c,d) $tt$ pair production, and (e,f) $tt\bar{u}$ 
production.
 \label{feyn}}
\end{figure}

Various models of new physics (NP) have been invoked to explain the size of the asymmetry, with new physics appearing in the direct ($s$) channel or exchange ($t$) channel~\cite{AfbNP,Cao:2010zb}.   A model based on the exchange of a  non-universal massive neutral vector boson $Z^\prime$ is sketched in Fig.~\ref{feyn}(b).  It is intriguing because it naturally produces top quarks in the forward region of rapidity via the process $u\bar{u}\to t\bar{t}$, with a $Z^\prime$ in the $t$-channel.  This approach requires a flavor changing neutral current (FCNC) interaction $u$-$t$-$Z^\prime$, 
\begin{equation}
\mathcal{L}= g_W \bar{u}\gamma^\mu (f_L P_L + f_R P_R) t Z^\prime_\mu + h.c., 
\end{equation}
where $g_W$ denotes the weak coupling strength.  Since the left-handed coupling $f_L$ is highly 
constrained by $B_d$-$\bar{B}_d$ mixing: $f_L<3.5\times 10^{-4}~ (m_{Z^\prime}/100~{\rm GeV})$~\cite{Cao:2010zb}, it is set to zero ($f_L=0$) for the remainder of this discussion.  The coupling of the 
$Z^\prime$ to charm quarks $Z^{\prime}-u-c$ is also set to zero since, otherwise, there would be a tree level contribution to $D$-$\bar{D}$ mixing.   

Figure~\ref{feyn}(a) displays the dominant leading-order QCD SM production of a 
$t\bar{t}$ pair at the Tevatron, while Fig.~\ref{feyn}(b) shows $Z^\prime$-induced 
$t\bar{t}$ pair production.  
A NP contribution to  $A_{\rm FB}$ arises from the absolute square of the NP contribution (Fig.~\ref{feyn}(b)) 
and the interference between the NP and the full set of NLO SM QCD amplitudes.  
To produce a large enough asymmetry, the coupling $f_R$ must be large if the $Z^\prime$ is heavy.  However,  the model is tightly constrained by other Tevatron data.  The coupling $f_R$ cannot be so large as to produce disagreement with the measured $t \bar{t}$ total cross section and the $t\bar{t}$ invariant mass distribution.  In our published paper~\cite{Berger:2011ua} we derive quantitative bounds on $f_R$ and $m_{Z^\prime}$ from Tevatron measurements of $A_{\rm FB}$ and the $t \bar{t}$ total cross section.   

Once $Z^\prime$ exchange is invoked as in Fig.~\ref{feyn}(b), then the processes shown in Figs.~\ref{feyn}(c) and~\ref{feyn}(d) predict that same-sign top quark pair production should exist.  The large valence quark $u u$ parton luminosity at the Large Hadron Collider (LHC) means that same-sign $tt$ pair production will be large.   In Ref.~\cite{Berger:2011ua}, we use the use the values we derive for  $f_R$ to predict the size of  the $t t$ pair production cross section at the LHC.  The CMS collaboration responded quickly to these predictions~\cite{Chatrchyan:2011dk}.  Their bounds on  same-sign $t t$ pair production at the LHC show that the simple $Z^\prime$ exchange explanation of the Tevatron asymmetry is disfavored.   

In Ref.~\cite{Berger:2011ua}, we focus on the collider phenomenology of  $tt$ pair production in early LHC experiments with 7~TeV center-of-mass (c.m.) energy and $1 {~\rm fb}^{-1}$ integrated luminosity.  In addition to predictions for the rate of same-sign $tt$ pairs,  
we show that the expected right-handed top quark polarization could be measured.  
We further consider same-sign $tt$ pair production in association with a jet, 
as depicted in Fig.~\ref{feyn}(e) and (f), from 
which one can obtain the invariant mass of the $Z^\prime$ from the reconstructed top 
quarks and the additional jet.   Note that there is no resonance in the $tt$ invariant 
mass spectrum since both top quarks are produced in the $t$-channel.   Our inclusive cross 
sections for $tt$ and $tt\bar{u}$ are shown in Ref.~\cite{Berger:2011ua} as a function of the 
$Z^\prime$ mass.   The $tt\bar{u}$ rate is smaller because it relies on the gluon-quark 
luminosity, smaller than the large valence $uu$ luminosity.   

%%%%%%%%%%%%%%%%%%%%%%%%%%%%%%%%%%
\section{Analysis of the Tevatron Data}
The forward-backward rapidity asymmetry $A_{\rm FB}$ is defined as  
\begin{eqnarray}
A_{\rm FB} & = &  \frac{\sigma_F-\sigma_B}{\sigma_F + \sigma_B} =
 \frac{\sigma_{F}^{\rm SM}-\sigma_{\rm B}^{\rm SM}+\sigma_{\rm F}^{\rm NP}-\sigma_{\rm B}^{\rm NP}}
      {\sigma_{F}^{\rm SM}+\sigma_{\rm B}^{\rm SM}+\sigma_{\rm F}^{\rm NP}+\sigma_{\rm B}^{\rm NP}}
       \label{eq:AFB1} \nonumber \\
 & = & 
 A_{\rm FB}^{\rm NP}\times R+A_{\rm FB}^{\rm SM}\left(1-R\right)
 \label{eq:AFB}
\end{eqnarray}
where $\sigma_{F(B)}$ denotes the $t\bar{t}$ cross section in the forward (F) and backward (B) rapidity region, and 
\begin{eqnarray}
A_{\rm FB}^{\rm NP} &\equiv&
(\sigma_{\rm F}^{\rm NP}-\sigma_{\rm B}^{\rm NP})/(\sigma_{\rm F}^{\rm NP}+\sigma_{\rm B}^{\rm NP}),
\nonumber \\
A_{\rm FB}^{\rm SM} &\equiv&
(\sigma_{\rm F}^{\rm SM}-\sigma_{\rm B}^{\rm SM})/(\sigma_{\rm F}^{\rm SM}+\sigma_{\rm B}^{\rm SM})
\nonumber \\
R&\equiv&(\sigma_{\rm tot}^{\rm NP})/(\sigma_{\rm tot}^{\rm SM}+\sigma_{\rm tot}^{\rm NP}) 
\label{eq:def}
\end{eqnarray}
are the asymmetries induced by NP and in the SM, and $R$ is the fraction of the NP contribution to the total cross section.  The standard model QCD and new physics contributions to the cross sections 
are denoted by superscripts SM and NP.  
\begin{figure}
\centering
\includegraphics[clip,scale=0.9]{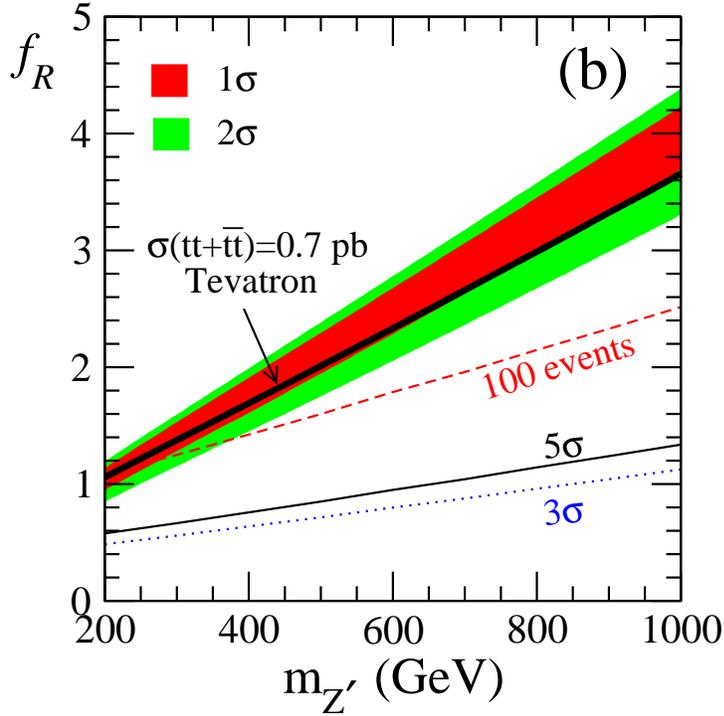}
\caption{
The shaded bands in the plane of $m_{Z^\prime}$ and $f_R$ are determined from the 
fit to $A_{\rm FB}$ and $\sigma(t\bar{t})$ in Ref.~\cite{Berger:2011ua}; the inner (outer) band corresponds to $1\sigma$ ($2\sigma$) C.L.  Lines are drawn for $5\sigma$ and $3\sigma$ discovery of same sign $tt$ pairs at 7~TeV with an integrated luminosity of $1~{\rm fb}^{-1}$, after all cuts are imposed, as specified in the text.  
A dashed line shows the expectation for 100 signal events.   The Tevatron limit on 
$f_R$ from the direct search for same-sign top quark pairs is presented.
\label{xsec}}
\end{figure}
There are tight constraints on the model from Tevatron data alone.   The inclusive cross section for $t \bar{t}$ production agrees with QCD SM expectations within the uncertainties of both experiment and theory, so the cross section itself limits the magnitude of the right-handed coupling $f_R$ from above, as do the data on the $t \bar{t}$ invariant mass spectrum.   On the other hand, the observed large size of  $A_{\rm FB}$ bounds $f_R$ from below.  The shaded regions in the $f_R$ plane in Fig.~\ref{xsec} are derived from requiring consistency with both $A_{\rm FB}$~\cite{Aaltonen:2011kc} and the $t\bar{t}$ production cross section $\sigma(t\bar{t})$~\cite{CDF:9913}:
\begin{eqnarray}
A_{\rm FB} & = & 0.475 \pm 0.114~~{\rm for}~~m_{t\bar{t}}\geq 450~{\rm GeV} \nonumber \\
\sigma(t\bar{t}) & = &7.50 \pm 0.48~{\rm pb}.
\end{eqnarray}
The inner (red) and outer (green) regions correspond to $1\sigma$ and $2\sigma$ C.L., respectively. 
The SM predictions of  $A_{\rm FB} (m_{t\bar{t}}\geq 450~{\rm GeV})$ and  $\sigma(t\bar{t})$ calculated with $m_t = 172.5~{\rm GeV}$ are $0.088$~\cite{Aaltonen:2011kc} and $6.9~{\rm pb}$~\cite{Cao:2010zb}, respectively.
The lower bound of each band is derived from the $A_{\rm FB}$ measurement 
while the upper bound is from the $\sigma(t\bar{t})$ data.  
\begin{figure}[ht].
\centering
\includegraphics[scale=0.45]{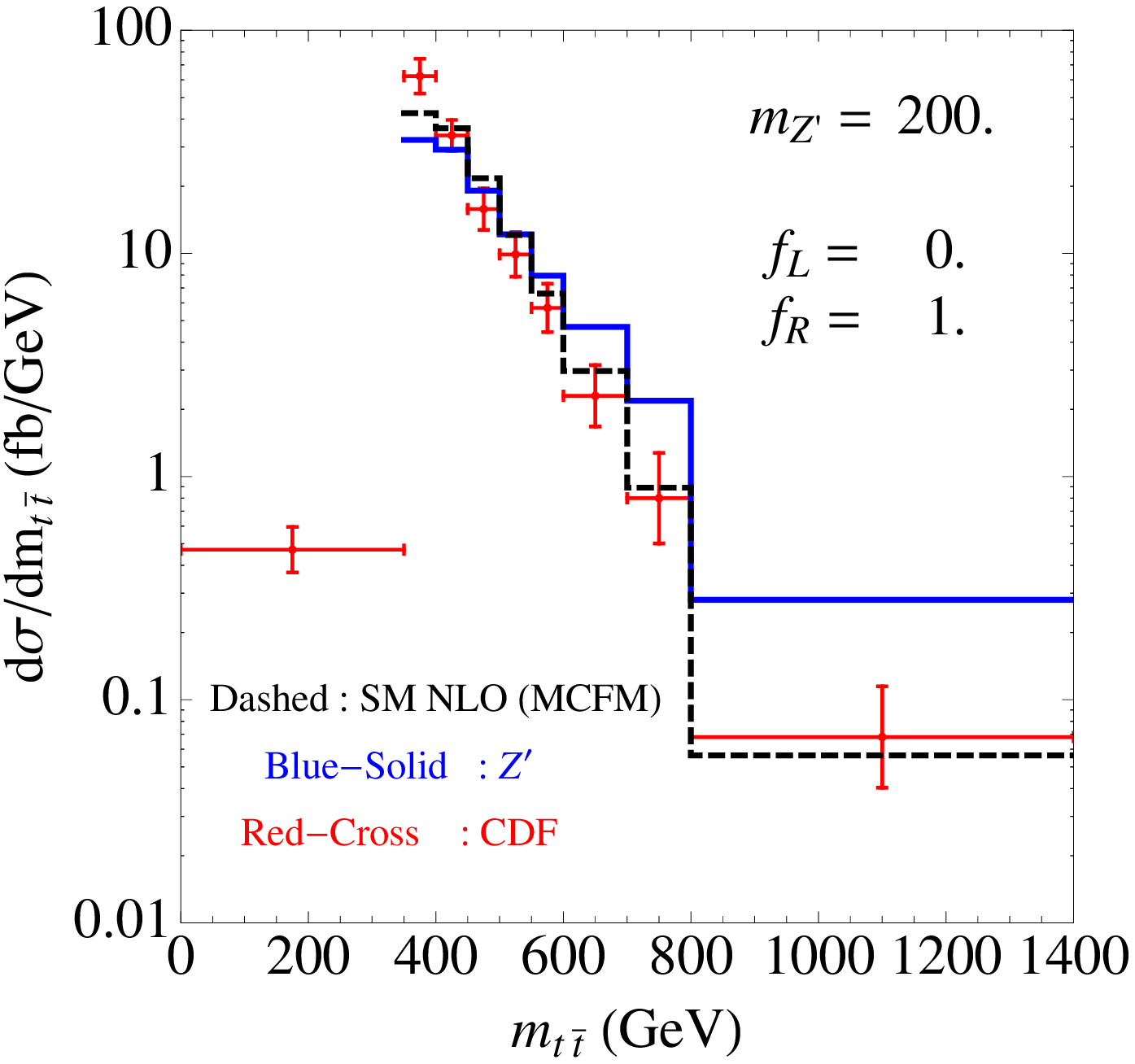}
\hspace{1cm}
\includegraphics[scale=0.45]{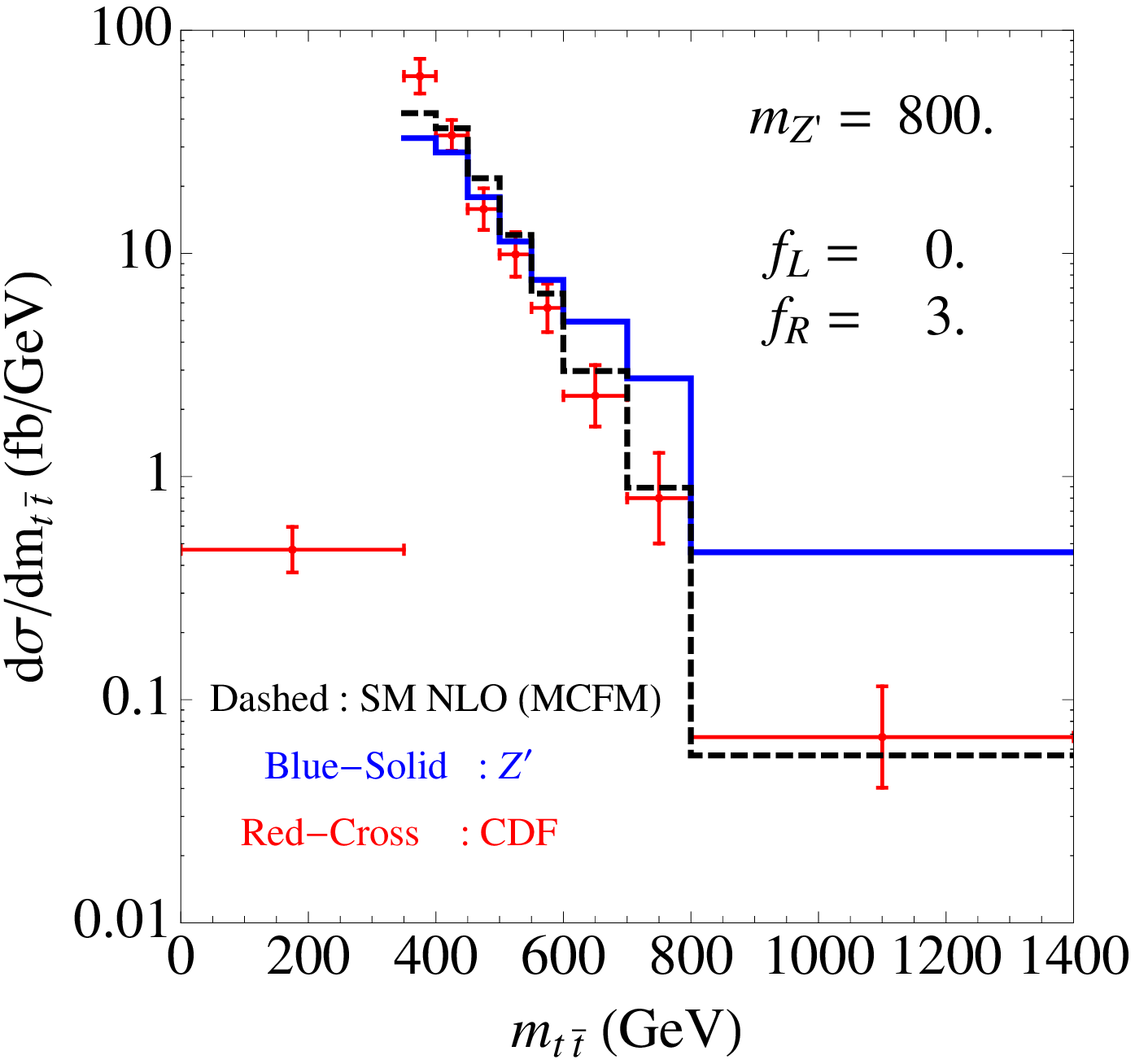}
\caption{Distribution in the invariant mass of the $t \bar{t}$ pair.  Shown are the CDF 
data~\cite{Aaltonen:2009iz} and the NLO QCD standard model contribution (dashed line).  Inclusion of 
the new physics $Z^{\prime}$ contribution results in the solid curve.  In the left plot the $Z^{\prime}$ mass is 200 GeV and $f_R = 1$, while these parameters are 800 GeV and $f_R = 3$ in the right plot.}
\label{massplots} 
\end{figure}  

In addition we verify that our computed distribution in the invariant mass $m_{t\bar{t}}$ is consistent with CDF data~\cite{Aaltonen:2009iz} at the level of $\lesssim 2\sigma$ deviations.   Examples of the effect of the $Z^{\prime}$-exchange contribution on the invariant mass distribution are shown in 
Fig.~\ref{massplots} for two sets of values of ($f_R$, $m_{Z^{\prime}}$).   The $Z^{\prime}$ exchange 
contribution depresses the distribution at small values of the $t \bar{t}$ invariant mass and increases it 
at large values.  The overall effect is a change in the slope of the invariant mass distribution.  It is worth noting in passing here that the the data in the bin at large mass, 800 to 1400 GeV,  are the most sensitive to the presence of an effect in the $s$ channel, such as a heavy $t \bar{t}$ resonance.      

Referring to Figs.~\ref{feyn}(c) and~\ref{feyn}(d), we see that same-sign top quark pair production 
should be present at the Tevatron, albeit with a rate that depends on the small $u$ quark density in an antiproton.  The search for same sign top quark pairs at the Tevatron, 
$\sigma(tt+\bar{t}\bar{t}) \lesssim 0.7~{\rm pb}$~\cite{Aaltonen:2008hx},  
imposes a constraint on $f_R$ and $m_{Z^{\prime}}$ shown by the black band in Fig.~\ref{xsec}(b). 
Parts of the otherwise allowed $1\sigma$ and $2\sigma$ bands are then excluded by these data.
If the cut acceptance were dropped by a factor of 2, to 0.25\%, then the same sign bound would increase to $\lesssim 1.4~{\rm pb}$, and the entire shaded region in Fig.~\ref{xsec} would be allowed.  

%%%%%%%%%%%%%%%%%%%%%%%%%%%%%%%%%%
\section{Predictions for the LHC}
In order to trigger on same-sign $tt$ events at the LHC, we demand that 
both top quarks decay leptonically and we further concentrate on the $\mu^+$ as its charge
can be better determined~\cite{Aad:2009wy}.   The sample of events of interest to us is defined by 
$\mu^+\mu^+~ b~b$ plus $\met$, 
where the missing transverse momentum $\met$ originates from two unobserved neutrinos.  
Our procedure for simulating the signal and background processes at the parton level, retaining all spin correlations, is similar to that described in Refs.~\cite{Berger:2010fy,Zhang:2010kr}.  The dominant SM backgrounds are:
\begin{eqnarray}
pp & \to & W^{+}(\to \ell^+\nu)W^{+}(\to \ell^+\nu)jj, \\
pp & \to & t\bar{t} \to bW^{+}(\to \ell^+\nu)\bar{b}(\to \ell^+)W^{-}(\to jj), 
\end{eqnarray}
computed with ALPGEN~\cite{Mangano:2002ea}.  
Other SM backgrounds, e.g. triple gauge boson production 
($WWW$, $ZWW$, and $WZg(\to b\bar{b})$), 
occur at a negligible rate after kinematic cuts.  
Since muon charge identification is not perfect,  we remark that  
$t\bar{t}$ pair production could also be a background when $\mu^-$ leptons from 
the antitop quark decay are misidentified as $\mu^+$ leptons. However,  
this background is negligible~\cite{Zhang:2010kr}.

At the analysis level, we require all signal and background events to pass the following acceptance cuts:
\begin{eqnarray}
&& n_j = 2,~n_{\mu^+} = 2,~p_T^j\geq 50\,{\rm GeV}, ~\left|\eta_{j}\right|\leq 2.5,~\nonumber \\
&& p_{T}^{\ell}\geq 50\,{\rm GeV},\quad \left|\eta_{\ell}\right|\leq2.0, \quad \met>20~{\rm GeV},\nonumber \\
&&\Delta R_{jj,j\ell,\ell\ell} > 0.4,
\label{eq:cut}
\end{eqnarray}
where the separation $\Delta R$ in the azimuthal angle ($\phi$)-pseudorapidity ($\eta$) plane between the objects $k$ and $l$ is
$\Delta R_{kl}\equiv\sqrt{\left(\eta_{k}-\eta_{l}\right)^{2}+\left(\phi_{k}-\phi_{l}\right)^{2}}$.
The two jets are further required to be $b$-tagged.
We also model detector resolution effects as described in Ref.~\cite{Zhang:2010kr}.
The SM backgrounds are suppressed efficiently such that less than 1 background event survives after cuts with an integrated luminosity of $1~{\rm fb}^{-1}$. 
Based on Poisson statistics, one needs 8 signal events 
in order to claim a $5\sigma$ discovery significance on top of 1 background event. 

The discovery potential is plotted in Fig.~\ref{xsec} with  black-solid  ($5\sigma$) and blue-dotted ($3\sigma$) curves.  
The values of $f_R$ indicated by the shaded bands in Fig.~\ref{xsec} show that $f_R \gtrsim 1$ for 
all $m_{Z^\prime} > 200$~GeV. They are everywhere above the 
values needed for $5$ standard deviation observation of same sign $t t$ pair production at the LHC.  
We conclude in Ref.~\cite{Berger:2011ua} that if {\it no} $tt$ signal is 
observed with 1~fb$^{-1}$ of integrated luminosity at the LHC, then simple non-universal 
$Z^\prime$ exchange cannot explain the Tevatron forward-backward asymmetry. 

\begin{figure}[ht]
\centering
\includegraphics[clip,scale=0.6]{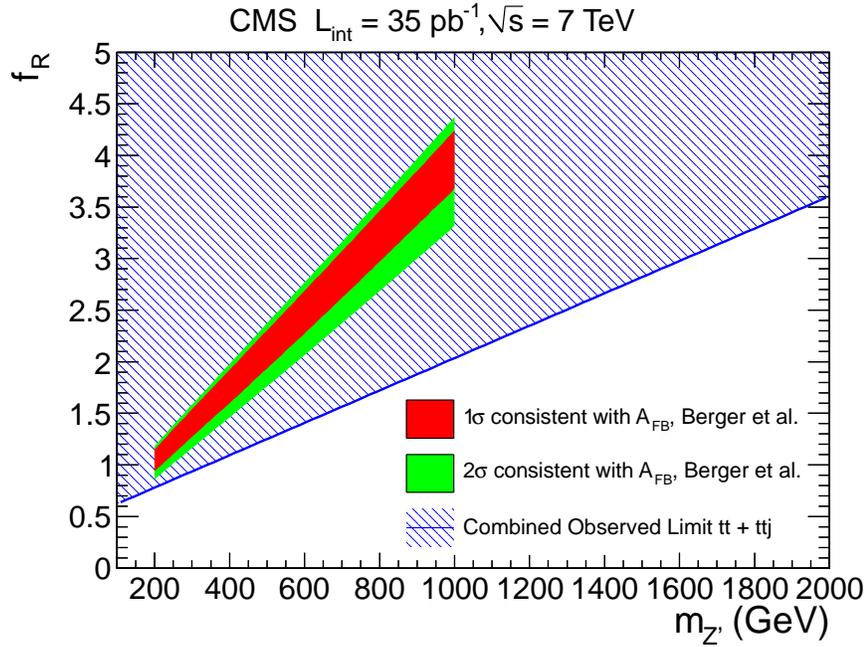}
\caption{The CMS exclusion region at 95\% CL is shown as a function of the $Z^{\prime}$
mass for various values of the right-handed coupling $f_R$, from Ref.~\cite{Chatrchyan:2011dk}.  The region above the solid line is excluded by the CMS data.    Also shown is the region of parameter space consistent wit the Tevatron measurements of $A_{\rm FB}$ and $\sigma(t \bar{t})$, from Ref.~\cite{Berger:2011ua}.  
%\label{bottomline}}
\label{cmsxsec}}
\end{figure}

The CMS collaboration responded quickly to these predictions~\cite{Chatrchyan:2011dk}.  They searched for same sign $t t$ production in the decay mode in which both tops decay leptonically, 
$t \rightarrow Wb \rightarrow \ell \nu b$, with a signature of two isolated leptons, two or more jets, and 
$\met$.  Working with a sample based on $35~{\rm pb}^{-1}$ of integrated luminosity, and imposing the kinematic cuts $p_T^{\ell} > 20$~GeV and $p_T^j  > 30$~GeV on the leptons and jets, they find 2 candidate events with a background expectation of $0.9 \pm 0.6$.   The bounds CMS obtains on  same-sign $t t$ pair production at the LHC are shown in Fig.~\ref{cmsxsec}.  Their results exclude the region in the ($f_R$, $m_{Z^{\prime}})$ plane that is consistent with $A_{\rm FB}$ at the Tevatron, meaning that the simple $Z^\prime$ exchange explanation of the Tevatron asymmetry is disfavored.   

%\begin{figure}[htb]
%

%%%%%%%%%%%%%%%%%%%%%%%%%%%%%%%%%%
\section{Top Quark Polarization}
\begin{figure}
\centering
\includegraphics[clip,scale=0.75]{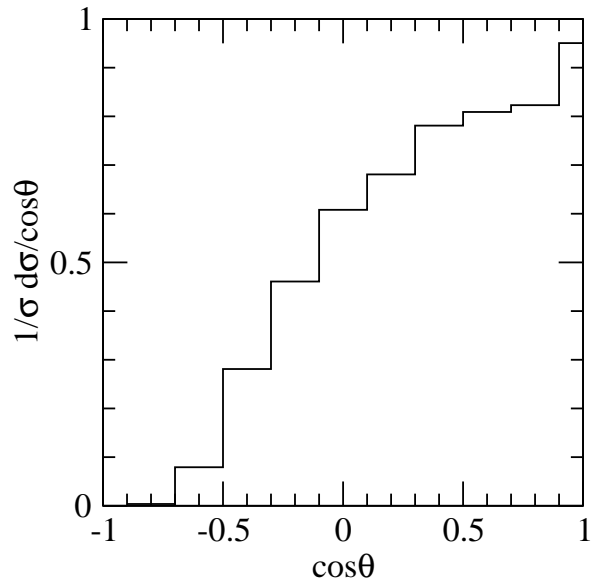}
\caption{Normalized distribution of the angle of the charged lepton relative to the top quark in the c.m. 
frame in $tt$ pair production after cuts and efficiencies are included, for the case  
$m_{Z^\prime}=800~\rm{GeV}$ and $f_R=1$.
\label{distro}}
\end{figure}
If an excess were observed in the $\mu^+\mu^+bb$ plus $\met$ sample,  one would seek to demonstrate consistency with a $uu\to tt$ origin or some other new physics source.  
Top quark polarization is a good probe of new physics.   For example, in the 
FCNC $Z^\prime$ model, the right-handed $u$-$t$-$Z^\prime$ coupling forces the top quarks 
to be mainly right-handed polarized.   Reconstruction of the two top quarks and measurement of their polarizations would permit validation of the model. 

Among the top quark decay products the charged lepton is maximally correlated with the top quark spin. 
In our signal process the charged lepton from  top quark decay exhibits a $1+\cos\theta$ distribution, 
where $\theta$ is the helicity angle between the charged lepton momentum in the top quark rest frame and top quark momentum in the c.m. frame of the production process. 
Following Ref.~\cite{Berger:2010fy}, we use the MT2 method~\cite{Lester:1999tx}
to select the correct $\mu$-$b$ combinations and to verify whether the final state is 
consistent with $t \to Wb$ parentage.  Then we make use of the on-shell
conditions of the two $W$ bosons and the two top quarks to solve for 
the neutrino momenta~\cite{Sonnenschein:2006ud,Bai:2008sk}.  
Once the neutrino momenta are known, the kinematics of the entire final state are fixed and 
the angular distribution may be constructed.

The reconstructed $\cos\theta$ distribution after cuts is plotted in Fig.~\ref{distro}, 
and it clearly shows the expected $1+\cos\theta$ form.
The discovery potential of the $tt\bar{u}$ signature is also promising.  If a peak can be 
found in the invariant mass spectrum of a $t$ and a  light jet 
(from the $\bar{u}$ in Fig.~\ref{feyn}(e) and (f)), 
one could confirm the presence of the FCNC $Z^\prime$.

\section{Summary}
Among possible new physics interpretations of the large forward-backward asymmetry observed in top quark production at the Tevatron~\cite{Aaltonen:2011kc,Abazov:2011rq}, the exchange of a flavor changing $Z^{\prime}$ is attractive in that it can generate a large asymmetry without structure in the $t \bar{t}$ invariant mass spectrum.  In the work~\cite{Berger:2011ua} summarized here, we fit the Tevatron asymmetry $A_{\rm FB}$ to extract the values of the two parameters of the model, the right handed coupling $f_R$ and the $Z^{\prime}$ mass $m_{Z^{\prime}}$.  
We verify that our fitted values are consistent with other Tevatron data including the distribution in the 
$t \bar{t}$ invariant mass and the bounds on the same-sign $tt$ cross section.  The values of the parameters are used to predict the expected rate of same-sign top quark pair production at the LHC, as a function of $f_R$ and $m_{Z^{\prime}}$.  A subsequent search for same-sign pairs by the CMS collaboration~\cite{Chatrchyan:2011dk} failed to find $tt$ events at the expected rate in a sample based on 35 pb$^{-1}$, excluding values of $f_R$ large enough to fit the Tevatron asymmetry for 
${m_{Z^{\prime}}} > 200$~GeV.  
We conclude that the simplest flavor changing $Z^{\prime}$ exchange model is not viable.  
Whatever physics eventually serves to explain the asymmetry, measurement of the polarization of top quarks is doable and would provide valuable added insight into the underlying dynamics.  

%%%%%%%%%%%%%%%%%%%%%%%%%%%%%%%%%%
\begin{acknowledgments}
I acknowledge many fruitful discussions with Qing-Hong Cao.  The research reported here was published in collaboration with Qing-Hong Cao, Chuan-Ren Chen, C.~S.~Li, and Hao Zhang.  It was supported in part by the U.S. DOE under Contract No.~DE-AC02-06CH11357.   
\end{acknowledgments}

\bigskip % extra skip inserted
%\begin{thebibliography}{9}   % Use for  1-9  references

\end{document}